\documentclass[iop]{emulateapj}
\usepackage{natbib}
\usepackage{graphicx}
\usepackage{times}
\usepackage{xspace} 
\usepackage{booktabs} 

\slugcomment{\it Draft 3.0 --- March 10, 2014}
\shorttitle{X-Ray Observations of WISE-Selected AGN}
\shortauthors{Stern et al.}

\def\ie{{i.e.}}
\def\eg{{e.g.}}

\def\wise{{\it WISE}}

\def\chandra{{\it Chandra}}
\def\xmm{{\it XMM-Newton}}
\def\spitzer{{\it Spitzer}}
\def\nustar{{\it NuSTAR}}

\def\wone{WISE~J1814+3412}
\def\wtwo{WISE~J2207+1939}
\def\wthree{WISE~J2357+0328}

\def\deg{\ifmmode {^{\circ}}\else {$^\circ$}\fi}
\def\kms{\ifmmode {\rm\,km\,s^{-1}}\else
    ${\rm\,km\,s^{-1}}$\fi}
\def\ergcm2s{\ifmmode {\rm\,erg\,cm^{-2}\,s^{-1}}\else
    ${\rm\,erg\,cm^{-2}\,s^{-1}}$\fi}
\def\ergAcm2s{\ifmmode {\rm\,erg\,cm^{-2}\,s^{-1}\,\AA^{-1}}\else
    ${\rm\,erg\,cm^{-2}\,s^{-1}\,\AA^{-1}}$\fi}
\def\ergs{\ifmmode {\rm\,erg\,s^{-1}}\else
    ${\rm\,erg\,s^{-1}}$\fi}
\def\kmsMpc{\ifmmode {\rm\,km\,s^{-1}\,Mpc^{-1}}\else
    ${\rm\,km\,s^{-1}\,Mpc^{-1}}$\fi}

\def\spose#1{\hbox to 0pt{#1\hss}}
\def\simlt{\mathrel{\spose{\lower 3pt\hbox{$\mathchar"218$}}
     \raise 2.0pt\hbox{$\mathchar"13C$}}}
\def\simgt{\mathrel{\spose{\lower 3pt\hbox{$\mathchar"218$}}
     \raise 2.0pt\hbox{$\mathchar"13E$}}}

\def\plotfiddle#1#2#3#4#5#6#7{\centering \leavevmode
\vbox to#2{\rule{0pt}{#2}}
\includegraphics{#1}}

\begin{document}

\title{NuSTAR and XMM-Newton Observations of Luminous,\\ Heavily
Obscured, WISE-Selected Quasars at $z \sim 2$}

\author{D.~Stern\altaffilmark{1}, 
G.~B.~Lansbury\altaffilmark{2}, 
R.~J.~Assef\altaffilmark{3},
W.~N.~Brandt\altaffilmark{4,5},
% ALPHABETICAL:
D.~M.~Alexander\altaffilmark{2},
D.~R.~Ballantyne\altaffilmark{6},
M.~Balokovi\'c\altaffilmark{7},
D.~Benford\altaffilmark{8},
A.~Blain\altaffilmark{9},
S.~E.~Boggs\altaffilmark{10},
C.~Bridge\altaffilmark{7},
M.~Brightman\altaffilmark{11},
F.~E.~Christensen\altaffilmark{12},
A.~Comastri\altaffilmark{13},
W.~W.~Craig\altaffilmark{10,14},
A.~Del~Moro\altaffilmark{2},
P.~R.~M.~Eisenhardt\altaffilmark{1},
P.~Gandhi\altaffilmark{2},
R.~Griffith\altaffilmark{4},
C.~J.~Hailey\altaffilmark{15},
F.~A.~Harrison\altaffilmark{7},
R.~C.~Hickox\altaffilmark{16},
T.~H.~Jarrett\altaffilmark{17},
M.~Koss\altaffilmark{18},
S.~Lake\altaffilmark{19},
S.~M.~LaMassa\altaffilmark{20},
B.~Luo\altaffilmark{4,5},
% S.A.~Stanford\altaffilmark{XXX}, -- CONFIRM
C.-W.~Tsai\altaffilmark{1},
D.~J.~Walton\altaffilmark{7},
E.~L.~Wright\altaffilmark{19},
J.~Wu\altaffilmark{19}, 
L.~Yan\altaffilmark{21} and
W.~W.~Zhang\altaffilmark{8}}

\altaffiltext{1}{Jet Propulsion Laboratory, California Institute
of Technology, 4800 Oak Grove Drive, Mail Stop 169-221, Pasadena,
CA 91109, USA [e-mail: {\tt daniel.k.stern@jpl.nasa.gov}]}

\altaffiltext{2}{Department of Physics, University of Durham, South
Road, Durham DH1~3LE, UK}

\altaffiltext{3}{N\'ucleo de Astronom\'ia de la Facultad de
Ingenier\'ia, Universidad Diego Portales, Av. Ej\'ercito Libertador
441, Santiago, Chile}

\altaffiltext{4}{Department of Astronomy and Astrophysics, The
Pennsylvania State University, 525 Davey Lab, University Park, PA
16802, USA}

\altaffiltext{5}{Institute for Gravitation and the Cosmos, The
Pennsylvania State University, University Park, PA 16802, USA}

\altaffiltext{6}{Center for Relativistic Astrophysics, School of
Physics, Georgia Institute of Technology, Atlanta, GA 30332, USA}

\altaffiltext{7}{Cahill Center for Astronomy and Astrophysics,
California Institute of Technology, Pasadena, CA 91125, USA}

\altaffiltext{8}{NASA Goddard Space Flight Center, Greenbelt, MD
20771, USA}

\altaffiltext{9}{Physics \& Astronomy, University of Leicester, 1
University Road, Leicester, LE1 7RH, UK}

\altaffiltext{10}{Space Sciences Laboratory, University of California,
Berkeley, 7 Gauss Way, Berkeley, CA 94720-7450, USA}

\altaffiltext{11}{Max-Planck-Institut f\"ur extraterrestrische
Physik, Giessenbachstrasse 1, D-85748, Garching bei M\"unchen,
Germany}

\altaffiltext{12}{Danish Technical University, DK-2800 Lyngby,
Denmark}

\altaffiltext{13}{INAF Osservatorio Astronomico di Bologna, via
Ranzani 1, I-40127, Bologna, Italy}

\altaffiltext{14}{Lawrence Livermore National Laboratory, Livermore,
CA 94550, USA}

\altaffiltext{15}{Columbia Astrophysics Laboratory, Columbia
University, New York, NY 10027, USA}

\altaffiltext{16}{Department of Physics and Astronomy, Dartmouth
College, 6127 Wilder Laboratory, Hanover, NH 03755, USA}

\altaffiltext{17}{Astrophysics, Cosmology and Gravity Centre,
Department of Astronomy, University of Cape Town, Rondebosch, South
Africa}

\altaffiltext{18}{Institute for Astronomy, Department of Physics,
ETH Zurich, Wolfgang-Pauli-Strasse 27, CH-8093 Zurich, Switzerland}

\altaffiltext{19}{Division of Astronomy \& Astrophysics, University
of California, Los Angeles, Los Angeles, CA 90095-1547, USA}

\altaffiltext{20}{Department of Physics and Yale Center for Astronomy
and Astrophysics, Yale University, New Haven, CT 06520-8120, USA}

\altaffiltext{21}{Infrared Processing and Analysis Center, Department
of Astronomy, California Institute of Technology, Pasadena, CA
91125, USA}

\begin{abstract} 

% original (longer) abstract:
% We report on a pilot \nustar\ and \xmm\ program that has observed
% a small sample of extreme \wise-selected AGN at $z \sim 2$.  The
% parent sample, selected to be faint or undetected in the \wise\ 3.4
% ($W1$) and $4.6 \mu$m ($W2$) bands but bright at 12 ($W3$) and 22
% ($W4$) $\mu$m, are extremely rare, with only $\sim 1000$ so-called
% ``$W1W2$-dropouts'' across the extragalactic sky.  Optical spectroscopy
% reveals typical redshifts of $z \sim 2$, implying rest-frame mid-IR
% luminosities of $\nu L_\nu(6 \mu{\rm m}) \sim 6 \times 10^{46}\,
% {\rm erg}\, {\rm s}^{-1}$ and bolometric luminosities that can
% exceed $L_{\rm bol} \sim 10^{14}\, L_\odot$.  The corresponding
% {\it intrinsic} hard X-ray luminosities are $L(2-10\, {\rm keV})
% \sim 4 \times 10^{45}\, {\rm erg}\, {\rm s}^{-1}$ for typical quasar
% templates.

We report on a \nustar\ and \xmm\ program that has observed
a sample of three extremely luminous, heavily obscured \wise-selected
AGN at $z \sim 2$ in a broad X-ray band ($0.1 - 79$~keV).  The parent sample, selected to be faint or
undetected in the \wise\ $3.4 \mu$m ($W1$) and $4.6 \mu$m ($W2$)
bands but bright at $12 \mu$m ($W3$) and $22 \mu$m ($W4$), are
extremely rare, with only $\sim 1000$ so-called ``$W1W2$-dropouts''
across the extragalactic sky.  Optical spectroscopy reveals typical
redshifts of $z \sim 2$ for this population, implying rest-frame
mid-IR luminosities of $\nu L_\nu(6 \mu{\rm m}) \sim 6 \times
10^{46}\, {\rm erg}\, {\rm s}^{-1}$ and bolometric luminosities
that can exceed $L_{\rm bol} \sim 10^{14}\, L_\odot$.  The corresponding
intrinsic, unobscured hard X-ray luminosities are $L(2-10\, {\rm keV})
\sim 4 \times 10^{45}\, {\rm erg}\, {\rm s}^{-1}$ for typical quasar
templates.  These are amongst the most luminous AGN known, though
the optical spectra rarely show evidence of a broad-line region and
the selection criteria imply heavy obscuration even at rest-frame
$1.5\, \mu{\rm m}$.  We designed our X-ray observations to obtain
robust detections for gas column densities $N_{\rm H} \leq 10^{24}\,
{\rm cm}^{-2}$.  In fact, the sources prove to be fainter than these
predictions.
Two of the sources were observed by both \nustar\
and \xmm, with neither being detected by \nustar\ ($f_{\rm 3-24\
keV} \simlt 10^{-13}\, {\rm erg}\, {\rm cm}^{-2}\, {\rm s}^{-1}$),
and one being faintly detected by \xmm\ ($f_{\rm 0.5-10\ keV} \sim
5 \times 10^{-15}\, {\rm erg}\, {\rm cm}^{-2}\, {\rm s}^{-1}$).  A
third source was observed only with \xmm, yielding a faint detection
($f_{\rm 0.5-10\ keV} \sim 7 \times 10^{-15}\, {\rm erg}\, {\rm
cm}^{-2}\, {\rm s}^{-1}$).  The X-ray data require gas column
densities $N_{\rm H} \simgt 10^{24}\, {\rm cm}^{-2}$, implying the
sources are extremely obscured, consistent with Compton-thick,
luminous quasars.  The discovery of a significant population of
heavily obscured, extremely luminous AGN does not conform to the
standard paradigm of a receding torus, in which more luminous quasars
are less likely to be obscured.  If a larger sample conforms with this finding, then this suggests an additional source
of obscuration for these extreme sources.

% Pulled from abstract:
% One of the sources considered here has type-2 AGN features in
% rest-frame ultraviolet spectroscopy obtained at Keck observatory,
% while the other two sources show Ly$\alpha$ emission with no clear
% AGN signatures, suggestive of even greater obscuring columns to
% their active nuclei.

\end{abstract}


\section{Introduction}

The {\it Wide-field Infrared Survey Explorer} (\wise) \citep{Wright:10}
is an extremely capable and efficient black hole finder.  As
demonstrated in selected fields by \spitzer\ \citep[\eg,][]{Stern:05,
Donley:12}, the same material that obscures AGN at UV, optical and
soft X-ray energies is heated by the AGN and emits strongly
at mid-IR wavelengths.  The all-sky \wise\, survey identifies
millions of obscured and unobscured quasars across the full sky
\citep[\eg,][]{Stern:12, Assef:13}, as well as very rare populations
of extremely luminous, heavily obscured AGN.

In terms of the latter, the \wise\, extragalactic team has been
pursuing sources that are faint or undetected in \wise\, $W1$
(3.4~$\mu$m) and $W2$ (4.6~$\mu$m), but are bright in $W3$ (12~$\mu$m)
and $W4$ (22~$\mu$m).  We refer to this population as $W1W2$-dropouts
\citep{Eisenhardt:12}.  This is a very rare population; selecting
to a depth of 1~mJy at 12~$\mu$m, there are only $\sim 1000$ such
sources across the extragalactic sky ($\sim 1$ per $30\ {\rm deg}^2$).
These objects are undetected by {\it ROSAT} and tend to be optically faint
($r \simgt 23$), below the detection threshold of SDSS.  We have
obtained spectroscopic redshifts for $> 100$ $W1W2$-dropouts thus
far, consistently finding redshifts $z \simgt 2$, with our current
highest redshift source at $z = 4.6$ \citep{Eisenhardt:14}.
Approximately half of the sources show clear type-2 AGN signatures
in the optical spectra, with the other half typically showing only Ly$\alpha$
emission, sometimes extended, which could be due to star formation
and/or AGN activity \citep{Bridge:13}.  The lack of a far-IR peak
in their broad-band SEDs suggests the dominant energy input for
this population comes from a heavily obscured AGN and not extreme
starbursts \citep[\eg,][]{Eisenhardt:12, Wu:12}.  Related high-luminosity
sources selected from the \wise\ satellite have also recently been
reported by \citet{Weedman:12} and \citet{Alexandroff:13}, while
several teams have identified less rare, less luminous sources from
\spitzer\ surveys with less extreme
colors \citep[\eg,][]{Dey:08, Fiore:09}.

Here we report on the first targeted X-ray follow-up of $W1W2$-dropouts.
We observed two sources with both the {\it Nuclear
Spectroscopic Telescope Array} \citep[\nustar;][]{Harrison:13} and
\xmm\ \citep{Jansen:01}; a third source was only observed by \xmm.
Unless otherwise specified, we use Vega magnitudes throughout and
adopt the concordance cosmology, $\Omega_M = 0.3$, $\Omega_\Lambda
= 0.7$ and $H_0 = 70\, \kmsMpc$.

% Our {\it Letter} is structured as follows:  \S2 briefly describes
% the X-ray sample, \S3 summarizes the multi-wavelength observations,
% and \S4 reports on the results, comparing the AGN luminosities and
% absorptions inferred from the mid-IR and the X-ray data.  We conclude
% in \S5.  

\section{Sample}

% w1814+3412 - (10may) + 10july; Keck/LRIS
% w2207+1939 - 10nov; Keck/LRIS
% w2357+0328 - 10nov; Keck/LRIS

\begin{figure}
% \plotone{3spectra.eps}
\plotfiddle{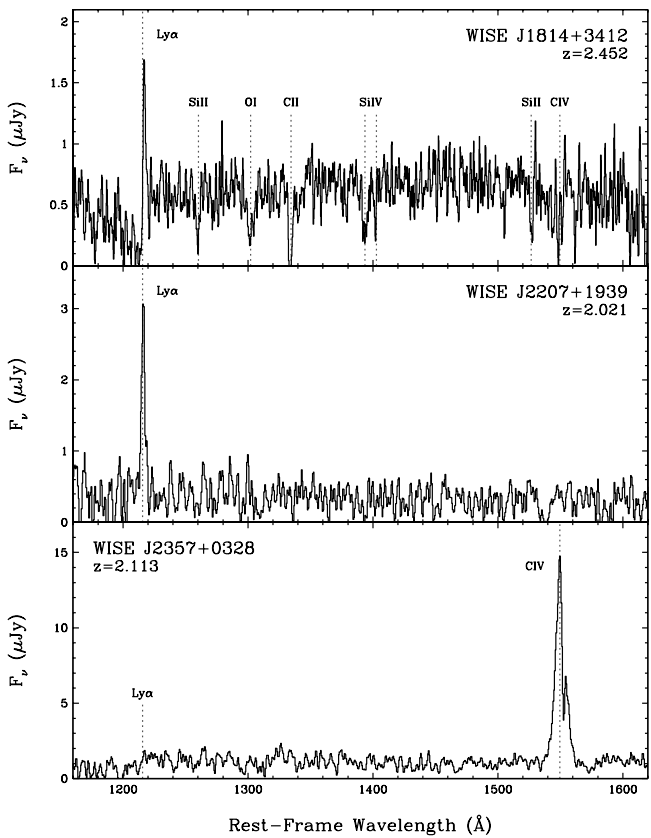}{4.4in}{0}{45}{45}{-133}{-20}
\caption{Keck/LRIS spectra of the three extreme \wise-selected
obscured AGN at $z \sim 2$ which we observed at X-ray energies.
\label{fig:spectra}}
\end{figure}

Fig.~\ref{fig:spectra} presents the optical spectra of the three
$W1W2$-dropouts targeted for X-ray follow-up:  WISEA~J181417.29+341224.8
(hereafter, \wone), WISEA~J220743.82+193940.1 (hereafter, \wtwo)
and WISEA~J235710.82+032802.8 (hereafter, \wthree).  The spectra
were all obtained with the Low Resolution Imaging Spectrometer
\citep[LRIS;][]{Oke:95} on the Keck~I telescope, between 2010 July
and 2010 November.  The sources were selected on the basis of having
unusually red colors across the \wise\ passbands in the initial
All-Sky \wise\ data release:  $W1 > 17.4$, $W4 < 7.7$, and $W2 -
W4 > 8.2$ \citep[for further details on the $W1W2$-dropout selection,
see][]{Eisenhardt:12}.  The most likely interpretation of sources
with these extreme colors is that they host an extremely luminous,
heavily obscured AGN which only becomes evident at observed wavelengths
$\simgt 10\mu$m.  The $W1$ flux limit essentially constrains the
sample to $z \simgt 1.5$ for the host galaxy not to be detected.

% Further details on the $W1W2$-dropout selection and spectroscopic
% follow-up are presented in \citet{Eisenhardt:14}.

Note the diversity of the optical spectra of the three sources
targeted for X-ray follow-up (Fig.~\ref{fig:spectra}).  This is
representative of the diverse optical spectroscopic properties of
the $W1W2$-dropout population in general \citep{Wu:12, Eisenhardt:14}.
\citet{Eisenhardt:12} discusses \wone\ in depth: briefly, the optical
spectrum is indistinguishable from an $L^*$ Lyman-break galaxy (LBG)
at $z \sim 2.5$ \citep[\eg,][]{Shapley:03}, with no obvious signature
of a (buried) AGN.  However, typical LBGs have $22 \mu$m flux
densities a factor of 1000 lower \citep{Reddy:06}.  The redshift
of \wtwo\ is based on a single, asymmetric, high equivalent width
emission line which is reliably identified as Ly$\alpha\ \lambda
1216$ \citep[for a detailed discussion of one-line redshifts,
see][]{Stern:00d}.  \wthree\ shows a high equivalent width ($\sim
300$~\AA, observed), slightly broadened (FWHM $\sim 1500\ {\rm km}\
{\rm s}^{-1}$), self-absorbed line identified as \ion{C}{4} $\lambda
1549$, which is a common strong line in AGN.  However, quite
unusually, no Ly$\alpha$ emission is evident.  The only strong
feature blue-ward of the emission line is a continuum break
at observed $\sim 3785$~\AA, which is consistent with the Ly$\alpha$
forest break for the longer wavelength emission line being \ion{C}{4}.
\citet{Hall:04} reports on a detailed investigation of a similar
SDSS quasar with broad \ion{C}{4} emission, but lacking broad Ly$\alpha$
emission.  They argue that the unusual spectrum cannot be solely
due to dust extinction in the broad line region (BLR), and instead
suggest that most, but not all, of the spectral properties can be
explained by an ununusually high density gas in the BLR ($n_{\rm
H} \sim 10^{15}\, {\rm cm}^{-3}$) with an incident power-law continuum
extending to $\geq 200 \mu$m.  Clearly the unusual optical spectrum
of \wthree\ is worthy of future study, but such analysis
is beyond the scope of the current paper which focuses on the X-ray
properties of the $W1W2$-dropout population.

In terms of their radio properties, \wone\ has a counterpart offset
by 6\farcs1 in the NRAO/VLA Sky Survey \citep[NVSS;][]{Condon:98}
with a flux density $S_{\rm 1.4\, GHz} = 3.4 \pm 0.5$~mJy.
\citet{Eisenhardt:12} report on follow-up radio observations of
this source with the Jansky Very Large Array which resolves the
NVSS emission into two distinct sources, the fainter of which is
associated with \wone.  Based on both its rest-frame 1.4~GHz radio
luminosity ($L_{\rm 1.4\, GHz} \sim 5 \times 10^{25}\, {\rm W}\,
{\rm Hz}^{-1}$) and its radio-to-optical ratio (rest-frame $L_{\rm
5\, GHz} / L_{\rm 0.44\, \mu m} \sim 200$), \wone\ qualifies as
radio-loud.  \wtwo\ has a counterpart offset by 7\farcs1 in the
NVSS with $S_{\rm 1.4\, GHz} = 5.2 \pm 0.4$~mJy, suggesting that
it is also radio loud.  \wthree\ has no radio counterpart in either
NVSS or the VLA Faint Images of the Radio Sky at Twenty Centimeters
survey \citep[FIRST;][]{Becker:95}.  The radio luminosities of the
first two sources further indicate the presence of a powerful AGN
in this \wise-selected population.

%
% FIGURE 1 - SPECTRA

%
% TABLE 1 - OBSERVING LOG
\begin{table*}[]
\centering
\caption{X-ray Observation Log.}
\begin{tabular}{lccccccc} \hline\hline \noalign{\smallskip}
\multicolumn{2}{l}{} & \multicolumn{3}{c}{\nustar} &
\multicolumn{3}{c}{\xmm} \\
\noalign{\smallskip}
\cmidrule(rl){3-5} \cmidrule(rl){6-8}
\noalign{\smallskip}
\multicolumn{1}{c}{} & $z$ & Observation ID & UT Date & Exposure & 
Observation ID & UT Date & MOS1/MOS2/pn Exposure \\
\multicolumn{1}{c}{Target Name} &  &  &  & (ks) &  &  & (ks) \\
\multicolumn{1}{c}{(1)} & (2) & (3) & (4) & (5) & (6) & (7) & (8) \\
\noalign{\smallskip} \hline \noalign{\smallskip}
\wone   & 2.452 & 60001114002 & 2012 Oct 30 & 21.3 & 0693750101 & 2012 Oct 07 & 29.6 / 29.6 / 19.6 \\
\wtwo   & 2.021 & 60001115002 & 2012 Oct 30 & 20.8 & 0693750201 & 2012 Nov 22 & 18.4 / 18.7 / 11.5 \\
\wthree & 2.113 & \nodata & \nodata & \nodata & 0693750401 & 2013 Jan 02 & 29.1 / 28.7 / 14.4 \\
\noalign{\smallskip} \hline \noalign{\smallskip}
\end{tabular}
\begin{minipage}[l]{0.94\textwidth}
\footnotesize
{\bf Note.} --- (1): Target name; full name and coordinates are in
Table~2.  (2): Redshift.  (3) and (4): \nustar\ observation ID and
start date.  (5): Net on-axis \nustar\ exposure time.  This value
applies for both FPMA and FPMB.  The target was on-axis for both
of the \nustar\ observations.  (6) and (7): \xmm\ observation ID
and start date.  (8): Net on-axis exposure time, corrected for
flaring and bad events, for the MOS cameras and the pn camera, as
indicated.
\end{minipage}
\label{table:obslog}
\end{table*}

\section{Observations and Multiwavelength Data}

\subsection{NuSTAR Observations}

In 2012 October \nustar\ obtained $\sim 20$~ks observations of
\wone\ and \wtwo; details of the observations, including net exposure
times, are provided in Table~\ref{table:obslog}.  We processed the
level~1 data using the \nustar\ Data Analysis Software (NuSTARDAS)
v.1.2.0, and produced calibrated and cleaned event files (level~2
data) for both \nustar\ focal plane modules (FPMA and FPMB) using
{\tt nupipeline} and the most current available version of the
Calibration Database files (CALDB 20130509).

\begin{figure}
\plotfiddle{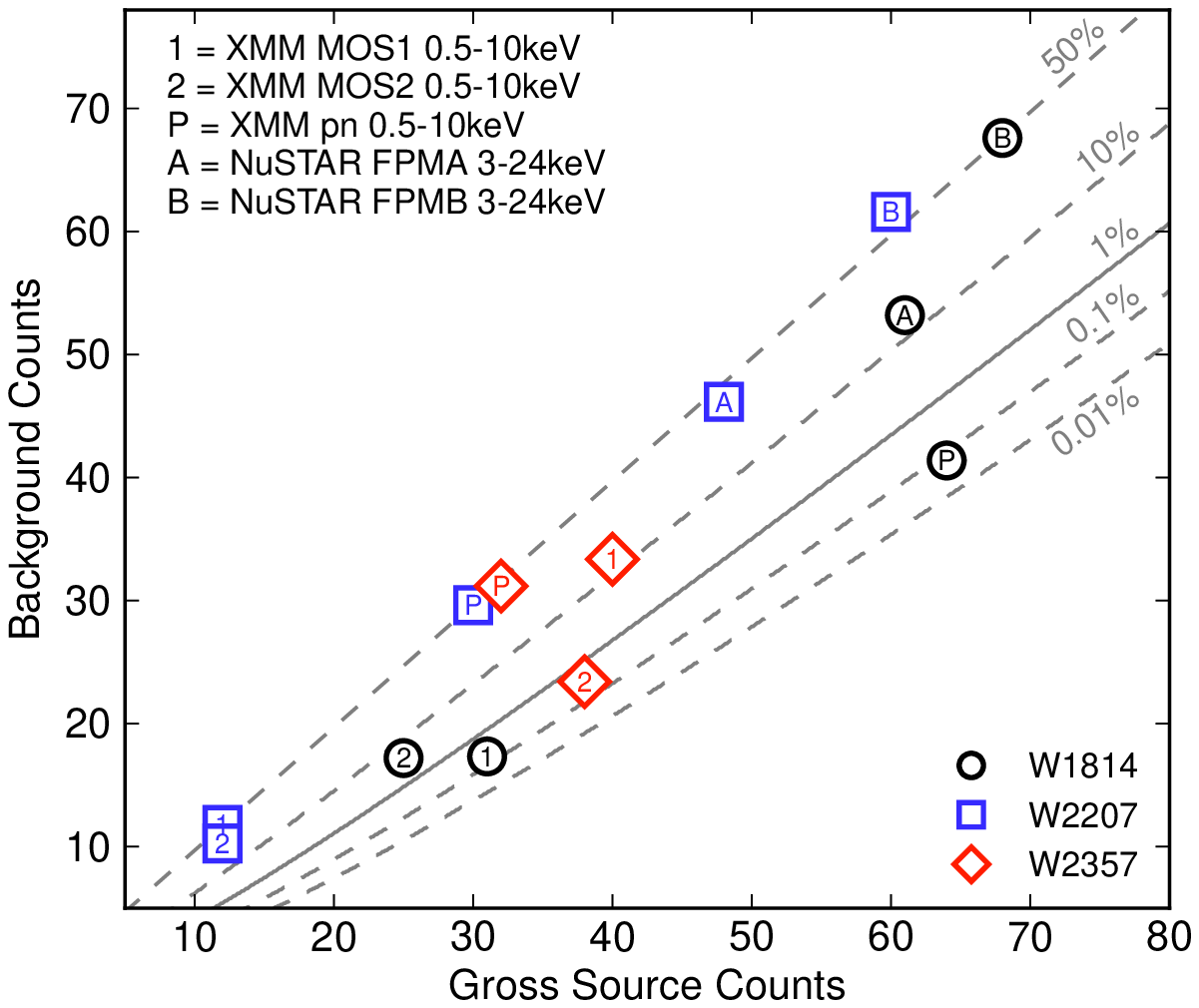}{2.8in}{0}{65}{65}{-190}{-150}
\caption{Gross source counts versus background counts (scaled to
the source region) for \wone, \wtwo\ and \wthree\ (circles, squares
and diamonds, respectively).  The 1, 2 and p labels correspond to
the 0.5-10~keV counts for the EPIC-MOS1, EPIC-MOS2 and EPIC-pn cameras
on \xmm, respectively.  The A and B labels correspond to the 3-24~keV
counts for FPMA and FPMB on \nustar, respectively.  The dashed lines
indicate Poisson no-source probabilities.  Two sources are faintly
detected: \wone\ is detected with the EPIC-MOS1 and EPIC-pn cameras,
and \wthree\ is detected with the EPIC-MOS2 camera.
\label{fig:Xraycounts}}
\end{figure}

Neither source was detected, though a serendipitous broad-lined AGN
at $z = 0.763$ was identified in the \wone\ field \citep{Alexander:13}.
We measured gross source counts in 45\arcsec\ radius apertures
centered on the \wise\ positions and local background counts from
an annulus of inner radius 90\arcsec\ and outer radius 150\arcsec\
centered on the sources.  We performed photometry in the observed-frame
3-24~keV, 3-8~keV and 8-24~keV bands, as well as the rest-frame
10-40~keV band for both FPMs and used binomial statistics to determine
the likelihood of the sources being detected.  Binomial statistics
are more accurate than Poisson statistics at these faint limits since
it takes into account uncertainty in the measured background (\ie,
it takes the total background counts into account, not just the
scaled background counts).  We use binomial statistics to
calculate the probability that the measured source counts are
purely due to background fluctuations \citep[\ie, false, or `no-source'
probabilities; for details, see][]{Lansbury:14}.  For both \wise\
sources the probability of a false \nustar\ detection based on the
binomial statistics is $> 15\%$.  As we take a no-source probability
$< 1\%$ to indicate a detection, neither source was detected.  Since
binomial statistics are not amenable to plotting simple tracks,
Fig.~\ref{fig:Xraycounts} shows the X-ray counts with Poisson
no-source probabilities.  The latter provide a good approximation
of binomial probabilities for our sources given the reasonably high
background count rates.

Table~\ref{table.phot} reports the 90\% confidence limit upper
limits to the flux in the 3-24~keV energy band, calculated using
the Bayesian method of \citet{Kraft:91}.  To convert count rate to
source flux, we used XSPEC v12.8.0k, taking into account the Response
Matrix File (RMF) and Ancillary Response File (ARF) for each FPM.
We assumed a power-law model with $\Gamma = 1.8$.  Assuming a harder
intrinsic spectrum does not qualitatively change these results;
\eg, adopting $\Gamma = 1.0$ only changes the rest-frame 10-40~keV
luminosities by $\leq 3\%$.

%
% FIGURE 2 - X-RAY BINOMIAL FALSE PROBABILITY

\subsection{XMM-Newton Observations}

We obtained $\sim 30$~ks observations of the three $W1W2$-dropouts
between 2012 October and 2013 January with the \xmm\ EPIC-MOS
\citep{Turner:01} and EPIC-pn \citep{Struder:01} cameras.  Details
of the observations are provided in Table~\ref{table:obslog}.  We
used data products from the Pipeline Processing System (PPS),
analyzed with the {\it Science Analysis Software}\footnote{\tt
http://xmm.esa.int/sas/} (SAS v.12.0.1).  We measured 0.5-10~keV
source counts in 15\arcsec\ radius apertures centered on the \wise\
positions.  For most observations, we measured local backgrounds
in slightly offset source-free circular apertures with radii of
$\sim 70 - 100\arcsec$ selected to avoid serendipitous sources and
chip gaps.  The exceptions were the EPIC-MOS observations of \wthree\
which had no nearby serendipitous sources and thus allowed for a
30\arcsec\ to 70\arcsec\ radius annulus centered the source position.
As above with the \nustar\ observations, we calculated binomial
false probabilities and plot the equivalent using Poisson statistics
in Fig.~\ref{fig:Xraycounts}.  Only \wone\ is reliably detected,
with a binomial no-source probability of 0.04\%\ with EPIC-pn and
0.1\%\ with EPIC-MOS1; \wone\ was not reliably detected with
EPIC-MOS2.  \wthree\ is weakly detected with EPIC-MOS2 with a
no-source probability of 0.2\%, and is undetected with the other
two cameras.  We used XSPEC to convert count rate (or count rate
limits) to flux, assuming a power-law model with $\Gamma = 1.8$,
and the \xmm\ RMFs and ARFs for each EPIC camera.  Again, adopting
a harder intrinsic X-ray spectrum would have a modest quantitative
effect on the derived luminosities, but would not affect our broad
conclusions.  Specifically, were we to instead assume $\Gamma =
1.0$, the rest-frame 2-10~keV luminosities would change by $\leq
18\%$.

% TMI on XMM:
% The \xmm\ spacecraft is equipped with three X-ray CCD cameras, which
% comprise the European Photon Imaging Camera (EPIC).  Two of the
% cameras are installed behind X-ray telescopes equipped with gratings
% that divert about half the incident light.  The remaining light
% reaches metal oxide semi-conductor (MOS) CCD arrays, referred to
% as the EPIC-MOS cameras \citep{Turner:01}.  The third CCD camera,
% using pn CCDs, has an unobstructed beam, and is referred to as the
% EPIC-pn camera \citep{Struder:01}.

\subsection{Mid-IR Data}

Table~\ref{table.phot} presents basic source properties and
multi-wavelength photometry for the three X-ray-targeted $W1W2$-dropouts.
We list mid-IR data from the AllWISE data release\footnote{\tt
http://wise2.ipac.caltech.edu/docs/release/allwise/} and \spitzer,
where the latter comes from the {\it Warm Spitzer} observations
reported by \citet{Griffith:12}.  Comparing the AGN luminosity and
reddening of \wone\ derived solely from the mid-IR photometry (\S
4.1) to the values in \citet{Eisenhardt:12} derived from 16-band
multi-wavelength photometry, we obtain consistent values within
$\sim 10\%$.

% Trimmed / dropped self references:
% Since our primary interest in this analysis
% is in the AGN, we focus solely on the mid-IR photometry here.
% Other investigations into the $W1W2$-dropout population focus on
% optical, sub-millimeter, and/or far-IR photometry which better
% constrain the host galaxy properties and cooler dust emission
% \citep[\eg,][]{Wu:12, Assef:14, Bridge:14, Jones:14, Tsai:14}.
% Specifically, comparing the AGN luminosity ...

%% TABLE 2 - SOURCE PROPERTIES
\begin{table*}[]
\centering
\caption{Source Properties.}
\begin{tabular}{cccc} \hline\hline \noalign{\smallskip}
& \wone & \wtwo & \wthree \\
\noalign{\smallskip} \hline \noalign{\smallskip}
R.A. (J2000)    	& 18:14:17.29	& 22:07:43.82	& 23:57:10.82 \\
Dec. (J2000)    	& +34:12:24.8	& +19:39:40.1	& +03:28:02.8 \\
$z$			& 2.452		& 2.021		& 2.113 \\
			&		&		& \\
$W1$ (3.4 $\mu$m)	& $18.861\pm0.440$ & $17.174\pm0.127$ & $>18.142$ \\
$W2$ (4.6 $\mu$m)	& $17.609\pm0.492$ & $16.136\pm0.170$ & $>16.614$ \\
$W3$ (12 $\mu$m)	& $10.410\pm0.061$ & $10.630\pm0.106$ & $10.088\pm0.068$ \\
$W4$ (22 $\mu$m)	& $ 6.863\pm0.071$ & $ 7.135\pm0.101$ & $ 6.942\pm0.112$ \\
$[3.6]$			& $17.707\pm0.023$ & $17.023\pm0.048$ & $17.487\pm0.076$ \\
$[4.5]$			& $17.021\pm0.020$ & $16.208\pm0.025$ & $16.544\pm0.036$ \\
% 			&		&		& \\
$f_{\rm 0.5-2\, keV}$   & $0.109\pm0.053$ & $<0.264$ 	& $<0.243$ \\
$f_{\rm 2-10\, keV}$    & $<1.30$         & $<0.780$ 	& $1.05\pm0.53$ \\ 
$f_{\rm 0.5-10\, keV}$  & $0.521\pm0.141$ & $<0.523$ 	& $0.73\pm0.31$ \\ 
$f_{\rm 3-24\, keV}$    & $<7.55$ 	  & $<6.04$ 	& \nodata \\
 		 	&		&		& \\
$L_{\rm 6\mu m}$    	& $20.10\pm2.40$& $8.28\pm1.62$ & $5.04\pm0.36$ \\
$E(B-V)_{\rm AGN}$  	& $15.1\pm1.1$  & $17.6\pm2.3$  & $5.5\pm0.4$ \\
$L_{\rm 2-10\, keV}$    & 1.04          & $<1.13$       & $<1.14$ \\
$L_{\rm 10-40\, keV}$   & $<17.1$       & $<10.1$       & \nodata \\
\noalign{\smallskip} \hline \noalign{\smallskip}
\end{tabular}
\begin{minipage}[l]{0.94\textwidth}
\footnotesize
{\bf Note.} --- Astrometry and \wise\ photometry are from the AllWISE
release; mid-IR photometry is all in Vega magnitudes and \wise\
limits report the 95\%\ confidence lower limit to the apparent
magnitude.  \spitzer\ photometry, in brackets, is from Griffith et
al. (2012).  X-ray fluxes, all in the observed frame, are in units
of $10^{-14} \ergcm2s$. $L_{\rm 6\mu m}$ is the rest-frame $6 \mu$m
luminosity ($\nu L_\nu$) of the AGN in units of $10^{46} \ergs$.
Rest-frame X-ray luminosities are in units of $10^{44} \ergs$.  We
report 68\% confidence limit (CL) uncertainties on the X-ray fluxes;
upper limits correspond to the 90\% CL.
\end{minipage}
\label{table.phot}
\end{table*}

\section{Results}

\subsection{AGN Properties from Mid-IR Data}

We modelled the spectral energy distribution (SED) of each source
using the \citet{Assef:10} 0.03-30 $\mu$m empirical AGN and galaxy
templates.  Each SED is modelled as a best-fit, non-negative
combination of an elliptical, spiral, and irregular galaxy component,
plus an AGN component.  Only the AGN component is fit for dust
reddening \citep[for further details on the SED models, see][]{Assef:08,
Assef:10}.  The modeling outputs $L_{\rm 6\mu m}$, the derived
intrinsic luminosity ($\nu L_\nu$) of the AGN component at rest-frame
$6 \mu$m, as well as the reddening of the AGN component, $E(B-V)_{\rm
AGN}$ (see Table~\ref{table.phot}).  For the typical gas to dust
ratio observed by \citet{Maiolino:01} for luminous AGN, the nuclear
reddening values of $E(B-V)_{\rm AGN} \sim 5 - 20$ imply gas columns
of $N_{\rm H} \sim (5 - 20) \times 10^{23}\ {\rm cm}^{-2}$, which
reach into the Compton-thick regime ($N_{\rm H} \geq 1.5 \times
10^{24}\ {\rm cm}^{-2}$).

%
% FIGURE 3 - Lx - L(6um)
\begin{figure*}
\plotfiddle{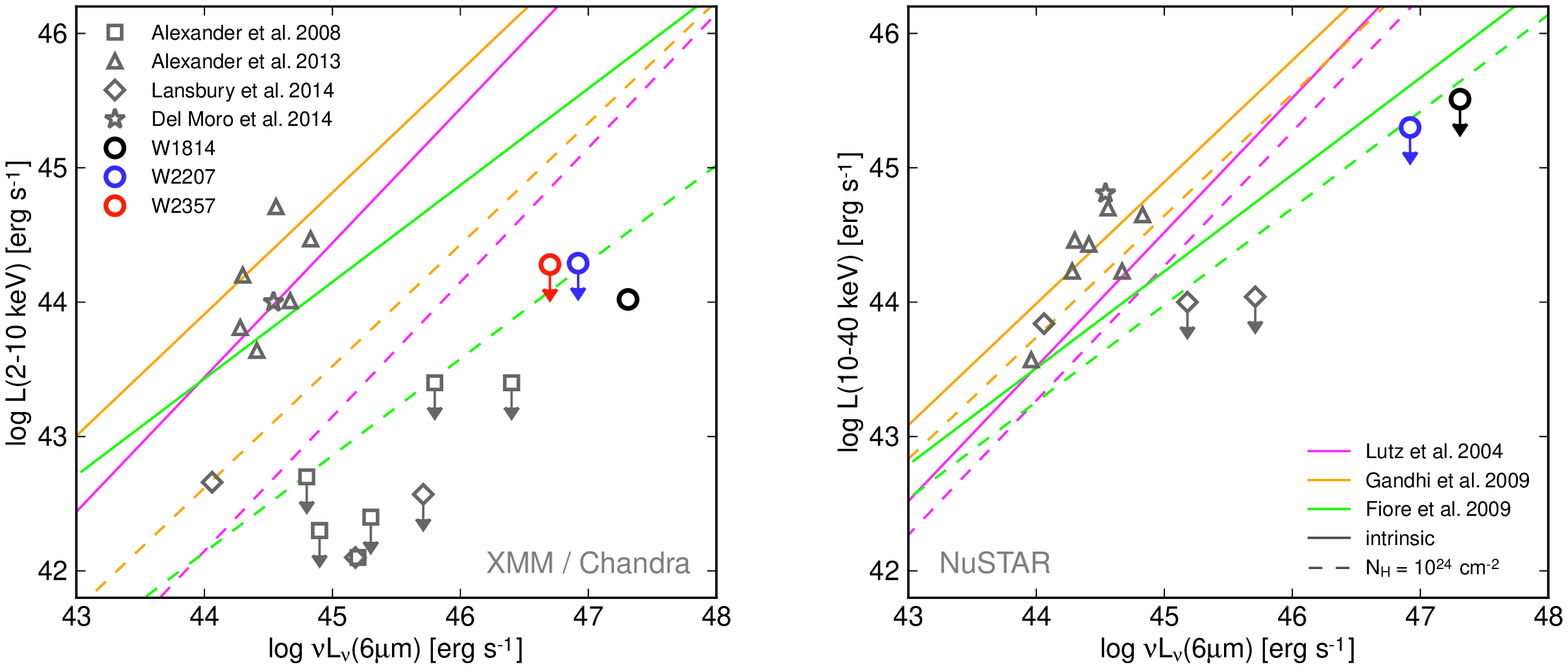}{2.7in}{0}{65}{65}{-190}{-150} 
\caption{Rest-frame X-ray luminosity against rest-frame $6 \mu$m
luminosity for: (left) 2-10 keV luminosities calculated using \xmm\
data; and (right) 10-40 keV luminosities calculated using \nustar\
data.  The X-ray luminosities are not corrected for absorption, and upper
limits correspond to $3\sigma$ values to aid literature comparisons.
\wone, \wtwo\ and \wthree\ are shown as black, blue and red circles,
as indicated.  We compare with \nustar\ observations from the
serendipitous survey (Alexander et al. 2013; triangles), a survey
of three SDSS type-2 quasars at $z \sim 0.5$ (Lansbury et al. 2014;
diamonds), and an interesting source at $z \approx 2$ detected by
\nustar\ in the ECDFS (Del~Moro et al. 2014; star); in the left
panel we also show soft X-ray data on Compton-thick quasars from
Alexander et al. (2008; squares).  We compare with three published
intrinsic relations for 2-10 keV calibrated using local AGN, as
indicated.  The relations are extrapolated to the 10-40 keV band
assuming $\Gamma = 1.8$ and in both panels the dashed lines show
the result of obscuration by $N_{\rm H} = 10^{24}\ {\rm cm}^{-2}$.
Assuming the low X-ray luminosities are due to absorption, sources
that lie below the $N_{\rm H} = 10^{24}\ {\rm cm}^{-2}$ tracks may
be Compton-thick.
\label{fig:LxLir}} 
\end{figure*}

\subsection{Indirect X-Ray Absorption Constraints}

The mid-IR properties of $W1W2$-dropouts indicate the presence of
extremely luminous, heavily obscured AGN, with bolometric luminosities
approaching, or even exceeding $L_{\rm bol} \sim 10^{14}\, L_\odot$
\citep{Eisenhardt:12, Wu:12, Bridge:13, Assef:14, Jones:14, Tsai:14}.
Given the heavy obscuration implied by their mid-IR SEDs and optical
spectra, it is perhaps unsurprising that the three sources we
targeted for X-ray follow-up are either undetected or only faintly
detected, despite observing at the penetrating energies above
rest-frame 10~keV which are less affected by absorption.  But what
are their absorption column densities?  We can obtain indirect
estimates of this from their mid-IR luminosities since unobscured
AGN tend to have a fairly tight relation between their mid-IR and
X-ray luminosities.

Fig.~\ref{fig:LxLir} compares the rest-frame $6 \mu$m and X-ray
luminosities for our targeted sources.  We show local relations
from \citet{Lutz:04}, \citet{Gandhi:09} and \citet{Fiore:09}, as
well as non-beamed AGN with $L_X > 10^{43} \ergs$ from the \nustar\
serendipitous survey \citep{Alexander:13}; the serendipitous sources
all lie within the scatter of the published relations.  We also
show Compton-thick quasars observed at soft energies from
\citet{Alexander:08}, three SDSS type-2 quasars observed by \nustar\
\citep{Lansbury:14}, and an obscured quasar at $z \approx 2$ detected
by \nustar\ in the Extended \chandra\ Deep Field South
\citep[ECDFS;][]{DelMoro:14}.  The literature obscured AGN and the
\wise-selected $W1W2$-dropouts generally lie significantly below
the published relations.  Assuming the suppression of their X-ray
emission is due to absorption rather than intrinsic X-ray weakness,
we can estimate their column densities from the dashed lines in
Fig.~\ref{fig:LxLir}, which apply columns of $N_{\rm H} = 10^{24}\
{\rm cm}^{-2}$ to the published relations.  These were calculated
using the MYTorus model \citep{Murphy:09} with photon index $\Gamma
= 1.8$ and a torus inclination angle $\theta_{\rm obs} = 70\deg$.
The implication is that all three targeted sources are obscured,
possibly heavily obscured or Compton-thick.  For the rest-frame
2-10~keV panel, the one detected source, \wone, has a column of
$N_{\rm H} \sim 10^{24}\, {\rm cm}^{-2}$ for the \citet{Fiore:09}
relation, and yet higher columns for the other two relations.  The
undetected sources require minimum columns of $N_{\rm H} \sim
10^{24}\, {\rm cm}^{-2}$ for all but the \citet{Fiore:09} relation.
Constraints from the \nustar\ nondetections are less strict, but
are again consistent with heavy absorption columns.  Furthermore,
two of the three \wise-selected targets are radio-loud.  Since
optically selected radio-loud quasars tend to have extra X-ray
emission quasars as compared to radio-quiet quasars \citep{Miller:11},
this suggests yet larger minimum absorption columns for \wone\ and
\wtwo.  In addition, the jet-linked X-ray emission is apparently
subject to similarly high obscuration as the nuclear X-ray emission.

% Finally, we note that \wthree\ shows somewhat broadened \ion{C}{4}~$\lambda
% 1549$ emission, which is not expected for a Compton-thick, true
% type-2 AGN.  On the other hand, this source has a much lower reddening
% inferred from the SED analysis (Table~\ref{table.phot}), consistent
% with the source being the least obscured of the three sources
% considered here.

\section{Discussion}

% We report on X-ray follow-up of three luminous ($L_{6 \mu{\rm m}}
% \simgt 10^{46}\ {\rm erg}\ {\rm s}^{-1}$) \wise-selected AGN with
% extremely red mid-IR colors.  The sources are at $z \sim 2$ and
% show evidence for significant absorption both from modeling their
% mid-IR SEDs and their optical spectra:  two sources show no evidence
% of a broad-line region while the third source shows somewhat
% broadened, self-absorbed \ion{C}{4}~$\lambda 1549$ emission with
% no Ly$\alpha$ emission.  

We designed our X-ray integration times to provide robust detections
for (i) typical intrinsic AGN SEDs, and (ii) gas column densities
$N_{\rm H} \simlt 10^{24}\, {\rm cm}^{-2}$.  None of the three
sources was strongly detected, implying that at least one of the
assumptions in our experimental design does not hold.

\nustar\ has now observed a range of obscured AGN, from famous,
local sources such as Mrk~231 \citep{Teng:14}, Circinus \citep{Arevalo:14},
NGC~424 \citep{Balokovic:14}, and NGC~4945 \citep{Puccetti:14}, to
higher redshift obscured quasars at $z \sim 0.5$ from SDSS
\citep{Lansbury:14} and $z \sim 2$ in the ECDFS \citep{DelMoro:14}.
A recurring theme of these observations is that several AGN which
are extremely luminous at certain wavelengths, such as in the mid-IR
or [\ion{O}{3}]~$\lambda 5007$, remain faint at X-ray energies.
For some objects, this is
the case even for the more penetrating hard X-rays
$> 10$~keV.  For some sources, such as optically bright ($B \simlt
16$) broad-absorption line (BAL) quasars from the Palomar-Green
(PG) survey \citep{Schmidt:83} and Mrk~231, we consider, and sometimes
even favor attributing the hard X-ray faintness to intrinsic X-ray
weakness \citep{Luo:13, Teng:14}.  Such intrinsic X-ray weakness
seems the most plausible scenario when the AGN appears unobscured
at certain wavelengths, through strong UV continuum emission, broad
emission lines, and/or weak X-ray spectra that are well described
by a moderately absorbed power-law.  For a more detailed description
of the intrinsically X-ray weak scenario, see \citet{Luo:13} and
\citet{Teng:14}.

% Intrinsic X-ray weakness could be an aspect of the ``disk-wind''
% model of BAL systems, in which the BAL wind is launched from the
% AGN accretion disk close to the black hole ($r \sim 0.01 - 0.1$~pc)
% and is radiatively driven by UV line pressure \citep[\eg,][]{Proga:00}.
% An intrinsically X-ray weak AGN could explain why the central engine
% does not over-ionize the gas, thereby quenching the line-driving
% mechanism.

For the three \wise-selected AGN discussed here, we instead favor
interpreting the X-ray faintness as being due to a typical AGN seen
through extremely high absorbing columns, consistent with their
mid-IR SEDs and optical spectra.  Indeed, the X-ray column constraints
from Fig.~\ref{fig:LxLir} are broadly consistent with the mid-IR
measurements given typical luminous AGN gas-to-dust ratios from
\citet{Maiolino:01}.  What is more surprising is that these sources,
amongst the bolometrically most luminous AGN known, appear heavily
obscured.  Various observations have shown that more luminous AGN
are less likely to be obscured \citep[\eg,][]{Ueda:03, Simpson:05,
Assef:13}.  This is consistent with the ``receding torus model'',
first proposed by \citet{Lawrence:91}, in which the height of the
torus is independent of luminosity while the the inner radius of
the torus, corresponding to the distance at which dust reaches its
sublimation temperature, increases with luminosity.  Therefore, in
this model, more luminous AGN have more sightlines into the nucleus
and thus have a lower likelihood of being obscured.  

The $W1W2$-dropout population is a rare population, with a surface
density of just one source per $\sim 30\, {\rm deg}^2$ in the
extragalactic sky.  The mid-IR luminosities imply intrinsic X-ray
luminosities of a few $\times 10^{45}\, {\rm erg}\, {\rm s}^{-1}$
from the relations of \citet{Lutz:04} and \citet{Gandhi:09}.
\citet{Just:07} report on X-ray follow-up of the most luminous
quasars in the SDSS available at the time, finding 34 quasars across
4188 deg$^2$, or one source per $\sim 120\, {\rm deg}^2$.  The X-ray
luminosities of this luminous quasar sample prove comparable
to the expected intrinsic X-ray luminosities from the \wise-selected
sample, implying the surprising discovery of comparable numbers of
obscured and unobscured quasars at the top of the luminosity function.
\citet{Assef:14} and \citet{Tsai:14} present more detailed comparisons
between the $W1W2$-dropout and luminous unobscured quasar populations.

The discovery of a significant population of heavily obscured,
extremely luminous AGN does not conform to the simple receding torus
model, suggesting an additional source of obscuration.  Indeed, the
models of \citet{Draper:10} predict that Compton-thick AGN should
be more common at higher redshift because of the high fueling rates
of quasars require significant gas reservoirs.  This is consistent
with models showing that mergers may be more prominent in fueling
AGNs at $z \sim 2$ \citep[\eg,][]{Hopkins:08, Draper:12}.  Indeed,
these models also predict that the unified model of AGN will break
down for high-luminosity AGNs at $z \simgt 1$ because the obscuration
is not confined to the nucleus \citep[see also][]{Draper:11}.

Further investigations into this interesting population are clearly
warranted.  Deeper X-ray observations, achieving robust detections
rather than faint and non-detections, would be valuable.  For
example, assuming the obscuration is from a nuclear torus, deeper
X-ray observations should detect narrow, reflected Fe K$\alpha$
fluorescent emission at rest-frame 6.4~keV \citep[\eg,][]{Nandra:07};
a significant non-detection could point towards obscuration on
larger scales than the torus.  Such large-scale obscuration could
also be probed by high-resolution imaging, such as far-IR observations
with ALMA, and, eventually, mid-IR observations with {\it JWST}.
Near-IR spectroscopy could also look for reddening in the AGN
narrow-line region \citep[\eg,][]{Brand:07}, which is expected to
be significantly larger than the torus.  A clearer understanding
of the geometry of the obscuring region combined with an improved
reckoning of the $W1W2$-dropout space density will enable us to
better place this population within the context of AGNs and galaxy
evolution.

\acknowledgements 
This work was supported under NASA Contract No. NNG08FD60C, and
made use of data from the {\it NuSTAR} mission, a project led by
the California Institute of Technology, managed by the Jet Propulsion
Laboratory, and funded by the National Aeronautics and Space
Administration. We thank the {\it NuSTAR} Operations, Software and
Calibration teams for support with the execution and analysis of
these observations.  This research has made use of the {\it NuSTAR}
Data Analysis Software (NuSTARDAS) jointly developed by the ASI
Science Data Center (ASDC, Italy) and the California Institute of
Technology (USA).  This publication makes use of data products from
the {\it Wide-field Infrared Survey Explorer}, which is a joint
project of the University of California, Los Angeles, and the Jet
Propulsion Laboratory/California Institute of Technology, funded
by the National Aeronautics and Space Administration.  We acknowledge
financial support from the Science and Technology Facilities Council
(STFC) grants ST/K501979/1 (GBL), ST/I001573/1 (DMA and ADM) and
ST/J003697/1 (PG).  RJA was supported by Gemini-CONICYT grant number
32120009.  RCH acknowledges support from NASA through ADAP award
NNX12AE38G and the National Science Foundation through grant number
1211096.

% or short version, if we're a Letter:
% This work was supported under NASA Contract No. NNG08FD60C, and
% made use of data from the \nustar\ mission, a project led by the
% California Institute of Technology, managed by the Jet Propulsion
% Laboratory, and funded by the National Aeronautics and Space
% Administration.

\smallskip
{\it Facilities:} \facility{Keck (LRIS)}, \facility{NuSTAR},
\facility{Spitzer (IRAC)}, \facility{XMM-Newton}, \facility{WISE}

\smallskip
\copyright 2014.  All rights reserved.

\clearpage
\end{document}